# From Consumption to Reflection

## Designing Human–AI Relations for Stable Reasoning

Rikard Rosenbacke*, Carl Rosenbacke[1], Victor Rosenbacke[1,2], Martin McKee[3]

[1]*Faculty of Medicine, Lund University, Sweden*
[2]*Department of Economics, Lund University School of Economics and Management, Sweden*
[3]*Department of Health Services Research and Policy, London School of Hygiene & Tropical Medicine, UK*
**Corresponding Author: rikard@rosenbacke.com*

## Abstract

Large language models (LLMs) have transformed how humans access information, but not how we reason with it. Their fluency accelerates consumption while bypassing the slow, reflective processes that underpin sound judgment. This paper introduces a new architecture, Relational Reflective Intelligence (RRI), an inference-time governance layer that operationalises reflection through auditable reasoning loops. RRI operates not inside the model but around it, providing a practical, inference-time structure for stable, auditable reasoning between humans and LLMs.

The core premise is that LLMs inherit the same cognitive vulnerabilities that shape human thought: intuitive shortcuts, map–territory confusion, and a preference for coherence over falsification. When humans and models share these tendencies, their errors compound. The resulting phenomenon (relational drift) emerges not from the model alone, but from the interaction between humans and machines. Addressing this requires an architectural shift. While the 2017 paper *Attention Is All You Need* gave machines the ability to model relations between words, this work argues that the next step is to model relations between AI output and human reasoning. The challenge is no longer prediction by correlation, but reasoning through collaboration.

RRI provides that missing layer. It integrates three components: the Rose-Frame, which diagnoses where reasoning is likely to fail; the Architect's Pen, which introduces short, targeted reflection steps precisely when risk arises; and the RRI workflow, which operationalises these steps during inference without retraining the model or slowing work. Together, they convert the human–AI dyad into a joint reasoning system with shared checkpoints, explicit conflict surfacing, and an auditable trail of assumptions.

The aim is not to make machines think like humans, nor to force humans to reason like machines, but to create a relational environment in which both can compensate for each other's weaknesses. RRI therefore reframes AI safety as a cognitive architecture problem: reliable decisions require structured reflection, and reflection must be built into the interaction itself.

## Structure of the Five-Paper Research Series

This paper is part of a five-paper research series examining how stable human–AI reasoning can be achieved by regulating not only model capabilities but also the human–AI relationship itself.

**Part I, *The Missing Knowledge Layer in AI: A Framework for Stable Human–AI Reasoning***[1]

establishes the system-wide gap and why it matters, motivating the need for a dedicated knowledge layer that governs human–AI reasoning across both human and model contributions, rather than focusing exclusively on model-level capabilities.

**Part II, *Beyond "Hallucinations": A Framework for Stable Human–AI Reasoning***[2] diagnoses the problem by analysing **why** human users systematically mistake linguistic fluency for epistemic understanding, and how large language models - trained on natural language - optimise for coherence and plausibility rather than factual correctness, leading to predictable forms of reasoning collapse.

**Part III, *Governing Reflective Human–AI Collaboration: A Framework for Epistemic Scaffolding and Traceable Reasoning***[3] proposes **what** must be governed: the human–AI relationship itself, introducing mechanisms for epistemic scaffolding, traceability, and alignment with emerging regulatory frameworks (e.g., the EU AI Act, WHO digital health standards, ISO guidelines).

**Part IV, *From Consumption to Reflection: Designing Human–AI Relations for Stable Reasoning***[4] shows **how** this relation can be implemented at the interface level, treating interaction design as the primary unit of analysis for stabilising reasoning during use.

**Part V, *Epistemic Control Loops in Large Language Models: An Architectural Proposal for Machine-Side Regulation***[5] provides a technical proposal for inference-time regulation within language models, including internal epistemic signals, control loops, actuation mechanisms, and memory. This component is currently under development and will be released in a subsequent preprint.

This paper corresponds to **Part IV** of the series and focuses on interface-level design to stabilise human–AI reasoning in use. It builds on the diagnosis in Part II and the governance framework in Part III by treating interaction design as the primary unit of analysis. System-wide framing is provided in Part I, while model-side regulation is developed separately in Part V.

# 1. Introduction

Large language models (LLMs) have become the most rapidly adopted cognitive tool in human history. They provide instant access to structured information, fluent text generation, and exploratory reasoning at unprecedented speed. Yet these advances have reshaped access far more than they have strengthened judgment. The fluency of model output encourages rapid consumption rather than slow evaluation. As answers become effortless, humans risk increasingly bypassing the deliberative steps that once structured reasoning. The core challenge is therefore cognitive rather than computational: when models produce instant coherence, people risk mistaking fluency for understanding.[2,3]

This tension reflects long-standing findings in cognitive science. Human reasoning is vulnerable to systematic distortions: intuitive impressions mimic understanding; representations are mistaken for reality; and coherence is preferred over falsification. These vulnerabilities, dominance of System-1 thinking, map, territory drift, and confirmation bias, are well documented in psychology and behavioural science.[2] LLMs now mirror and amplify these tendencies because they are trained on human language, which encodes the same shortcuts and illusions. When both humans and models share the same biases, their errors compound, producing a new category of cognitive risk: *relational drift*, where the interaction itself becomes the site of unnoticed error propagation.[3] At this point, three analytic layers begin to interact. At the cognitive level, we explain why human reasoning predictably fails when presented with fluent AI output. At the interaction level, we examine how those failures are amplified or mitigated by interface design. At the governance level, we ask how such interactions can be stabilised, documented, and regulated. The remainder of the paper moves explicitly across these layers, signalling which is in view at each step.

Current usage patterns heighten the urgency of this problem. LLMs are already employed in medicine, law, finance, education, and other regulated domains, often without documentation, auditability, or clear governance. Clinical reasoning, legal argumentation, and financial assessment are quietly shaped by model output despite the absence of evaluation or oversight standards. Unlike previous technologies that gradually altered cognition, LLMs immediately alter reasoning habits.[2,3] They make it easier to obtain answers than to interrogate them without structures that actively support reflection, thereby increasing the likelihood of drift toward superficial reasoning.

This concern aligns with recent frontier-lab framing that current AI systems are in a transitional “adolescence” phase:[6] highly capable, rapidly improving, yet still unreliable and socially isolated. Under this view, the most urgent requirement is not greater autonomy, but interaction-level stability: mechanisms that prevent fluent output from collapsing human reflection and that preserve accountable judgment under speed and scale.

**The Information Environment as a Cognitive Security Domain**

We now consider human–AI interaction as a security and accountability problem, rather than a purely cognitive or technical one. The NATO Strategic Communications Centre of Excellence (StratCom COE) argues that the emerging “Next Generation Information Environment” is defined by competition over interpretation, epistemic authority, and what counts as knowledge.[7] In this hybridised domain, private technology actors shape strategic outcomes, personalised influence systems erode institutional mediation, and adversarial manipulation, including data poisoning of open-source AI, makes epistemic reliability itself a contested resource. This framing strengthens the case for Relational Reflective Intelligence: when LLMs become epistemic mediators, governance cannot rely solely on model-level controls or institutional oversight, but must operate at the interaction layer, where interpretation is produced, and responsibility is either preserved or quietly dissolved.

The remainder of this paper series, therefore, treats the human–AI relation as the primary unit of analysis and develops a sequential architecture for restoring reflection, traceability, and epistemic stability under real-world conditions.

## Our Paper Trilogy: Why → What → How

The three middle papers in this series (Part II-IV) adopt a WHY → WHAT → HOW structure because this is the natural sequence by which knowledge, reasoning, and governance evolve (Table 1).[2–4] Scientific explanation begins by clarifying WHY a problem exists, then specifying WHAT conceptual structure is needed to address it, and finally determining HOW that structure can be operationalised. This pattern (purpose → form → mechanism) runs through cognitive science, systems theory, and engineering.

| Structural Question | Parts Title & Focus | RRI Component | Architectural Function |
|---|---|---|---|
| **WHY?** | Part II: *Beyond "Hallucinations": A Framework for Stable Human–AI Reasoning*[2] | The Rose-Frame | Diagnosis: Identifies the relational drift and cognitive traps that necessitate intervention. |
| **WHAT?** | Part III: *Governing Reflective Human–AI Collaboration: A Framework for Epistemic Scaffolding and Traceable Reasoning*[3] | The Architect's Pen | Definition: Presents the conceptual protocol for enforcing reasoning, deliberation and creating mutual intelligibility (Relational Intelligence). |
| **HOW?** | Part IV: From Consumption to Reflection: Designing Human–AI Relations for Stable Reasoning | The RRI Layer | Deployment: Operationalises reflection through the inference-time workflow, creating the Auditable Reasoning Trace, the traceable result. |

*Table 1: The Human-Side Trilogy (Parts II–IV)*

This paper presents the HOW: the full Relational Reflective Intelligence (RRI) architecture for auditable, reflective human–AI reasoning. Two companion works establish the foundation: the WHY, a diagnosis of epistemic drift in human–AI dyads (*Beyond "Hallucinations": A Framework for Stable Human–AI Reasoning*[2]); and the WHAT, the reflective primitive that governs reasoning (*Governing Reflective Human–AI Collaboration: A Framework for Epistemic Scaffolding and Traceable Reasoning*[3]). Readers focused solely on the deployable system may proceed directly; those seeking conceptual grounding may consult the earlier Parts II-III.

This triadic logic echoes foundational principles in cognitive science, as articulated by Herbert Simon, whose work on problem-solving and bounded rationality shaped both cognitive theory and early AI research. He argued that complex problem-solving requires a clear distinction among goals (WHY), problem representations (WHAT), and implementation procedures (HOW).[8] Modern AI leaders extend this logic. Demis Hassabis emphasises that advances in intelligence require clarity about the computational purposes of natural cognition (WHY) before designing architectures (WHAT) or learning regimes (HOW).[9] Yoshua Bengio similarly stresses that safe AI depends on identifying core cognitive requirements and "principled inductive biases" (WHY) to guide architectural choices (WHAT) and learning dynamics (HOW).[10] Yann LeCun highlights the same triad, distinguishing between an agent's objective (WHY), its world model and architecture (WHAT), and its learning process (HOW), noting that misalignment in any of these compromises reasoning.[11]

## Motivation for this work

Despite rapid progress in AI development, five critical gaps remain unaddressed across research, governance, and practice. These gaps motivate and necessitate the framework presented in this paper.

i) LLMs risk weakening human reasoning: A growing body of evidence suggests that frequent LLM use may shift users toward faster, less reflective decision-making. By presenting fluent answers instantly, LLMs reduce the natural friction that previously required humans to think through problems.[12] This creates cultural and cognitive vulnerability: when fluency is misinterpreted as truth, intuitive System 1 thinking predominates by default. Without explicit structures for selective reflection, human reasoning becomes more dependent on coherence and less grounded in evaluation.

ii) LLMs are already used in high-stakes regulated domains, often invisibly: Clinicians, lawyers, financial analysts, and educators routinely employ LLMs in ways that directly influence decisions. Yet most of this use remains in the shadows of existing governance frameworks. No audit trails capture how model outputs shape reasoning, and no standardised structures guide human evaluation. This creates unmonitored risk at both institutional and societal levels. Intervention is needed now, before informal practices become entrenched.

iii) Current AI-safety efforts focus almost exclusively on the machine: Most research aims to detect or mitigate hallucinations, quantify uncertainty, or build guardrails around model output.[3] These approaches improve the computational system but do not address the cognitive system formed by human + model. Reasoning does not occur solely within the model; it emerges from interaction. By focusing primarily on the machine, current frameworks overlook the relational layer in which drift, misunderstanding, and misplaced confidence arise.

iv) Emerging AI capabilities increase the need for structured reflective attention: As AI systems become more capable, discussions about alignment and "goal drift" intensify. But focusing only on preventing unwanted machine behaviour misses the foundational issue: humans themselves need structures that support slow thinking, falsification, and the exploration of conflict. Without reflective scaffolds, human–machine collaboration becomes increasingly vulnerable to unnoticed drift. Reflection is not an optional enhancement; it is the core safety mechanism for joint reasoning[3] so there is a need to embed governance externally, ensuring that reflection occurs in the interaction rather than inside the model.

v) Regulation now demands traceable, explicable human reasoning: The EU AI Act,[13] ISO 42001,[14] and responsible AI standards all require transparency, traceability, and documented human oversight for high-risk AI systems. But current LLM interaction patterns produce no auditable record of reasoning. Without a way to capture how humans reflect on model outputs, organisations cannot meet regulatory expectations. A system that restores reflection and records reasoning loops is therefore both a cognitive solution and a compliance mechanism.[3]

Where the landmark 2017 paper, *Attention Is All You Need*[15] taught machines to model relations between words, this paper argues that the next step is to model relations between human reasoning and machine output. The shift is from computational attention to reflective cognition. Attention built the model's fluency; reflection must make the human–AI system's reliability.

This paper, therefore, offers a practical, theoretically grounded, and regulatorily aligned method for transforming LLM use from passive consumption to active reflection, enabling humans and AI to think together with clarity, stability, and accountability.

## 2. Cognitive Science Foundations: Why Reasoning Needs Structure

Understanding why LLMs often reinforce cognitive shortcuts rather than correct them requires examining the historical foundations of reasoning research. Over the past seventy years, cognitive science has progressed through several major paradigms, each illuminating part of human thought, but none fully explaining how reasoning works when distributed across humans and machines. The missing element is the relational layer. Without it, AI safety strategies that focus solely on the model cannot address the cognitive vulnerabilities shaping human–AI collaboration.

The evolution of cognitive science can be traced through four traditions: (i) Cognitive Psychology (mind as symbolic processor), (ii) Neuroscientific Embodiment (mind as brain–body system), (iii) Embodied & Situated Cognition (mind as enacted through action), and (iv) Relational Cognition (mind as shared meaning-making between agents). LLMs expose the limits of the first three, making the fourth unavoidable.

**Cognitive Psychology: Mind as Internal Computation**

The symbolic revolution led by Herbert Simon and colleagues framed thinking as rule-based symbol manipulation, as captured in their work on human problem-solving.[16] Fodor extended this with the *Language of Thought* hypothesis, arguing that cognition operates over syntactic mental representations.[17] This architecture shaped early AI: intelligence was assumed to reside within a single system that performed internal computations. LLMs inherit this legacy as large pattern processors that generate coherent text but remain detached from grounded experience.

**Neuroscientific Embodiment: Thought as Sensorimotor Process**

Neuroscientific work by Damasio and others reframed cognition as deeply grounded in the body: perception, action, and affect shape reasoning.[18] More recent frameworks, such as Friston's predictive processing and the free energy principle, model the brain as an active inference engine that constantly updates expectations based on sensorimotor feedback.[19] In AI, this lens informs reinforcement learning and robotics, where intelligence emerges from interaction with an environment. Yet reasoning still resides within the organism; the unit of cognition is the single brain. LLMs, without bodies, cannot benefit from this kind of grounding.

**Embodied & Situated Cognition: Thought as Enacted**

Embodied and situated cognition was further developed, showing that humans think with and through their bodies, spatial representations, and tools. Lakoff and Johnson argued that bodily metaphors structure abstract concepts;[20] Tversky demonstrated how diagrams, spatial layouts, and gestures externalise thought[21]; Noë described perception as an active skill, enacted through sensorimotor engagement.[22] Cognition, on this view, is distributed across the brain, body, and environment. However, it still largely treats cognition as belonging to an individual agent interacting with the world, rather than to multiple reasoning systems coordinating with one another.

**The Missing Layer: Relational Cognition**

Relational traditions, most clearly anticipated by Vygotsky, Mead, and later Hutchins, treat thinking as inherently social and distributed. Vygotsky showed that higher mental functions are internalised from social interaction.[23] Mead analysed the self as emerging from taking the perspective of others;[24] Hutchins demonstrated that navigation and decision-making in real settings are properties of systems of humans and artefacts, not isolated minds.[25] This perspective aligns with the reality of scientific reasoning, legal argument, financial judgment, and clinical decision-making: they are fundamentally collaborative, negotiated, and corrected through dialogue.

When LLMs enter this landscape, they become additional agents or systems in the reasoning system. They do not think as humans do, but they participate in meaning-making by articulating patterns, generating counterfactuals, and rapidly reformulating ideas. This leads to our central claim:

*Reasoning in human–AI systems is relational. It emerges from coordination, correction, and reflection among systems (or agents), not from any one of them independently.*

Without explicit structures for reflection, human–AI interaction is prone to relational drift: humans and models reinforce the same intuitive, ungrounded, or overly coherent interpretations, creating mutually reinforcing error loops.

What follows shifts from explanation to diagnosis. Having established *why* cognition fails at the human–AI boundary, we now introduce a diagnostic framework that detects when and how interactional reasoning begins to drift. Here, the task is not to make machines autonomous thinkers, but to design interactions that:

- selectively engage System 2 reasoning,
- surface conflict rather than collapse prematurely into coherence,
- distinguish representation from reality,
- make reasoning explicit and inspectable, and
- detect and correct drift.

This requires a shift from purely computational architecture to *relational architecture*. The question becomes not "*How do we make the model reason better?*" but "*How do we structure the interaction so that humans and machines reason better together?*" The remainder of this paper develops exactly this structure: the Rose-Frame (diagnosis), the Architect's Pen (mechanism), and Relational Reflective Intelligence (inference-time implementation). We begin with the Rose-Frame, which operationalises these principles by detecting reasoning drift in human–AI interaction.

**Why LLMs Count as Partners in a Cognitive System**

Although LLMs lack embodiment, goals, or intentional mental states, they nonetheless participate in human reasoning in ways that fit core principles of relational and distributed cognition. Relational cognition, from Vygotsky to Hutchins, holds that higher-order thinking emerges not from isolated minds but from coordination between agents and artefacts engaged in a shared task.[25,26] In this tradition, tools that externalise, reorganise, or transform cognitive work function as participants in a wider cognitive system, even when they are not biological agents.

LLMs qualify as such participants because they shape reasoning in the same ways that cognitive artefacts do in classical distributed-cognition research. As Norman argues, artefacts that re-represent problems, expand working memory, or generate alternative frames become part of the functional architecture of thought.[27] LLMs perform all of these roles: they articulate patterns that humans cannot easily compute, externalise implicit assumptions, generate counterfactuals, and reorganise information in ways that directly influence how humans interpret a problem. Their contributions are therefore functionally agent-like, even without internal intentionality.

This functional agency is precisely what necessitates relational governance. When a tool begins to co-produce interpretations, shape epistemic framing, or collapse uncertainty prematurely through fluent output, it becomes part of the reasoning system rather than an external resource. The unit of analysis is no longer the model in isolation, but the human–LLM dyad, whose interaction can drift, converge prematurely, or reinforce shared biases. RRI responds to this shift by treating the interaction itself as the site of cognition and by providing an inference-time structure that stabilises the relational space in which reasoning emerges.

Under this view, LLMs do not need intentions to be cognitive partners; they need only to shape the dynamics of thought. The theoretical foundation for RRI, therefore, rests on established principles of distributed and relational cognition: cognitive stability depends not solely on the human or the model, but on the architecture of coordination between them.

## 3. The Rose-Frame: A Cognitive Architecture for Detecting Drift

Reasoning in human–AI systems emerges in the space between the two systems/agents. Our Part II paper[2] develops the Rose-Frame to diagnose where this interaction drifts away from grounded, reflective thought. It is essential for the present paper because it identifies when reasoning becomes unreliable and what type of drift the Architect's Pen undergoes in Part II[3] must correct. The Rose-Frame consists of three axes: Cognitive Mode (System 1–System 2, Epistemology–Ontology, and Conflict–Confirmation), each capturing a predictable failure pattern that arises when human intuition and model fluency reinforce one another.

The first axis, Cognitive Mode, concerns the balance between fast intuition (System 1) and slow deliberation (System 2).[28] Drift occurs when human intuition and model fluency combine to produce answers that feel correct without examination. A user accepts the first plausible answer, confuses coherence with correctness, skips falsification, or asks the model to confirm an initial hunch. The interaction becomes a loop of fluent but unexamined reasoning. The Rose-Frame identifies this moment, allowing the Architect's Pen to introduce a reflective step.

The second axis, Epistemology vs Ontology, addresses the confusion between representation and reality. Drift occurs when text is treated as evidence: a linguistic description is taken as a grounded fact, a narrative explanation is mistaken for a causal mechanism, or an abstract formulation is assumed to reflect real-world structure. LLMs amplify this because they simulate expertise rather than access it. When users treat textual completions as reality, reasoning becomes unmoored. The Rose-Frame flags these situations and re-anchors the interaction in observable, falsifiable facts.

The third axis, Conflict vs Confirmation, concerns how humans and LLMs handle uncertainty.[29] Drift happens when both collapse too quickly into the most coherent interpretation. Alternatives disappear, uncertainty is not expressed, and counterfactuals are left unexplored. The model smooths contradictions into a single narrative, while the user seeks answers rather than challenges. The result is convergence on coherence rather than accuracy. The Rose-Frame treats this premature agreement as a signal that reflective conflict is needed.

Although each axis captures a distinct cognitive vulnerability, they often reinforce one another: System 1 thinking amplifies ontology drift. Epistemic confusion encourages confirmation. Confirmation loops deepen System 1 dominance. Coherent representations mask the absence of grounding.

When all three align, human–AI drift becomes almost invisible. Answers appear correct, reasoning feels smooth, and both sides reinforce the same false assumptions. This is the danger zone for high-stakes decisions. For example: A physician using an AI diagnostic tool receives a confident recommendation (System 1 fluency), treats the probabilistic output as diagnostic certainty (map/territory collapse), and accepts it without checking contraindications (confirmation bias). All three axes fail simultaneously, yet the interaction feels smooth and professional. Without a mechanism to detect these patterns, errors propagate silently.

The Rose-Frame does not prescribe how to fix drift; that is the role of the Architect's Pen. Instead, it provides the diagnostic signals that determine when reflection must be initiated. It functions as a lightweight cognitive architecture that sits between the user and the model, guiding the interaction.

At inference time, the Rose-Frame can detect:

- high fluency with low evidence
- ambiguity in user intent
- rapid convergence without alternatives
- ontologically weak causal claims
- failure to distinguish patterns from phenomena

When these conditions appear, the system triggers a reflective step. In this way, the Rose-Frame acts as the cognitive radar for the entire interaction. It brings visibility to the invisible structure of drift. The following section explains how the Architect's Pen turns this diagnosis into a practical mechanism for structured reflection..

# 4. The Architect's Pen: Turning Reflection into a Cognitive Instrument

This section introduces a concrete interaction mechanism for restoring reflection once cognitive drift has been detected. The emphasis here is not on why reasoning fails, but on how reflective structure can be introduced at the interface level. If the Rose-Frame diagnoses where drift arises in the human–AI interaction, in our paper Part III[3], we describe the Architect's Pen that provides the mechanism for correcting it. The Pen is not a software tool in the narrow sense. It is a cognitive instrument, an interface layer designed to transform reflection from a moral aspiration into a practical, repeatable, auditable process. Its goal is straightforward: to ensure that when reasoning drifts, humans are guided back into selective, structured reasoning.

The Pen is built around a simple principle: reflection should occur only when it matters. It does not slow all interactions. It intervenes when the Rose-Frame detects heightened risk, when fluency dominates, grounding is weak, or coherence replaces conflict. In these moments, the Pen imposes structure, creating a short reflective loop that forces human reasoning to surface assumptions, test alternatives, and clarify intent.

The Architect's Pen acts as a regulator of joint cognition, not by altering internal model architecture, but by embedding reflective governance externally. This distinction ensures that reasoning is structured at the interaction layer, complementing rather than replacing architectural advances that remain simulations. It organises human–AI reasoning into a clear sequence and provides the scaffolding for reliable decision-making, especially in environments where the cost of error is high.

## The Three-Step Relational Loop

Before introducing the complete reflective-control cycle, the Pen operates through a simpler foundational structure: a three-step relational loop. This loop defines how humans and LLMs co-produce meaning. It has three moves:

**1. Human Framing (Intent and Abstraction)**

The human sets the initial direction, context, and level of abstraction. This step reveals intent, background knowledge, constraints, and what is considered relevant. Without explicit framing, LLMs guess intent using statistical inference, often inaccurately. Framing forces the human to articulate what they are actually trying to understand.

Examples: *"I need a differential diagnosis for abdominal pain with risk factors X and Y." "Summarise this contract only for liabilities affecting the seller." "Interpret this data through a Bayesian lens."*

Human framing is the first safeguard against drift, because it constrains the model's interpretive space.

**2. LLM Generation (Pattern Articulation)**

The model produces structured output: explanations, options, summaries, or hypothetical alternatives. At this stage, the LLM acts as an externalisation tool; it turns internal thoughts into visible representations, much like a pen helps an architect to visualise abstract ideas. Importantly, this step often surfaces patterns or alternatives that the human did not anticipate. The value is not in the correctness of the output, but in making reasoning visible**.**

**3. Human Reflection (Evaluation and Falsification)**

In the final move, the human assesses the output: *Which parts are grounded? Which patterns are hallucinated? What is missing? What conflicts or uncertainties remain?* This step is guided by the Rose-Frame, which highlights where reflection is necessary. It is also where the quality of reasoning is restored. Instead of consuming the model's output, the human evaluates it, searches for contradictions, and identifies gaps.

The loop then begins again. Each iteration becomes sharper, more grounded, and less vulnerable to drift. Depending on the complexity of the task, arriving at a grounded conclusion may require dozens, or even hundreds, of iterative passes. The relational loop establishes the architecture of collaboration.

## The Cognitive Basis of the Architect's Pen

Human reasoning is not composed of words. It unfolds through a continuous stream of spatial, sensory, and temporal impressions. We think by orienting ourselves in space, by comparing possibilities across time, by reacting to sensations of ease or tension, and by integrating fragments of perception that never fully reach language. Most reasoning is pre-verbal; words arrive late, often as a translation of patterns already taking shape in the mind.[21]

The Architect's Pen works because it intervenes at this pre-verbal level. It provides a momentary structure that helps the mind slow down, reorient, and externalise what would otherwise remain implicit. When the Pen prompts the user to state assumptions, consider alternatives, or revisit a judgment later, it is not asking for linguistic performance. It helps the mind reorganise its sensory field: to shift attention, to widen perceptual scope, to place the decision in a different temporal frame, or to engage with bodily cues that signal uncertainty.

In this sense, the Pen does what humans naturally struggle to do on their own. It stabilises attention when intuition accelerates too quickly. It anchors abstract reasoning in time and space by requiring the user to pause, imagine a counterfactual world, or picture how the decision would stand under scrutiny. And it uses language not as the medium of thought, but as a mirror for thought, a way to verbalise and clarify patterns that the mind already senses but cannot articulate without help.

Large language models amplify this process. Humans do not think in sentences, but LLMs excel at generating them. The Pen leverages this asymmetry: the model helps verbalise what the human senses, and the Pen ensures that this verbalisation does not become a substitute for reflection. Instead, the Pen coordinates both agents so that language becomes a tool for grounded reasoning rather than a fluent shortcut.

The following section integrates the Rose-Frame and Architect's Pen into a deployable workflow, Relational Reflective Intelligence (RRI), that operationalises reflection at inference time. The following section integrates these components into a full operational workflow.

# 5. Relational Reflective Intelligence: The Inference-Time Workflow

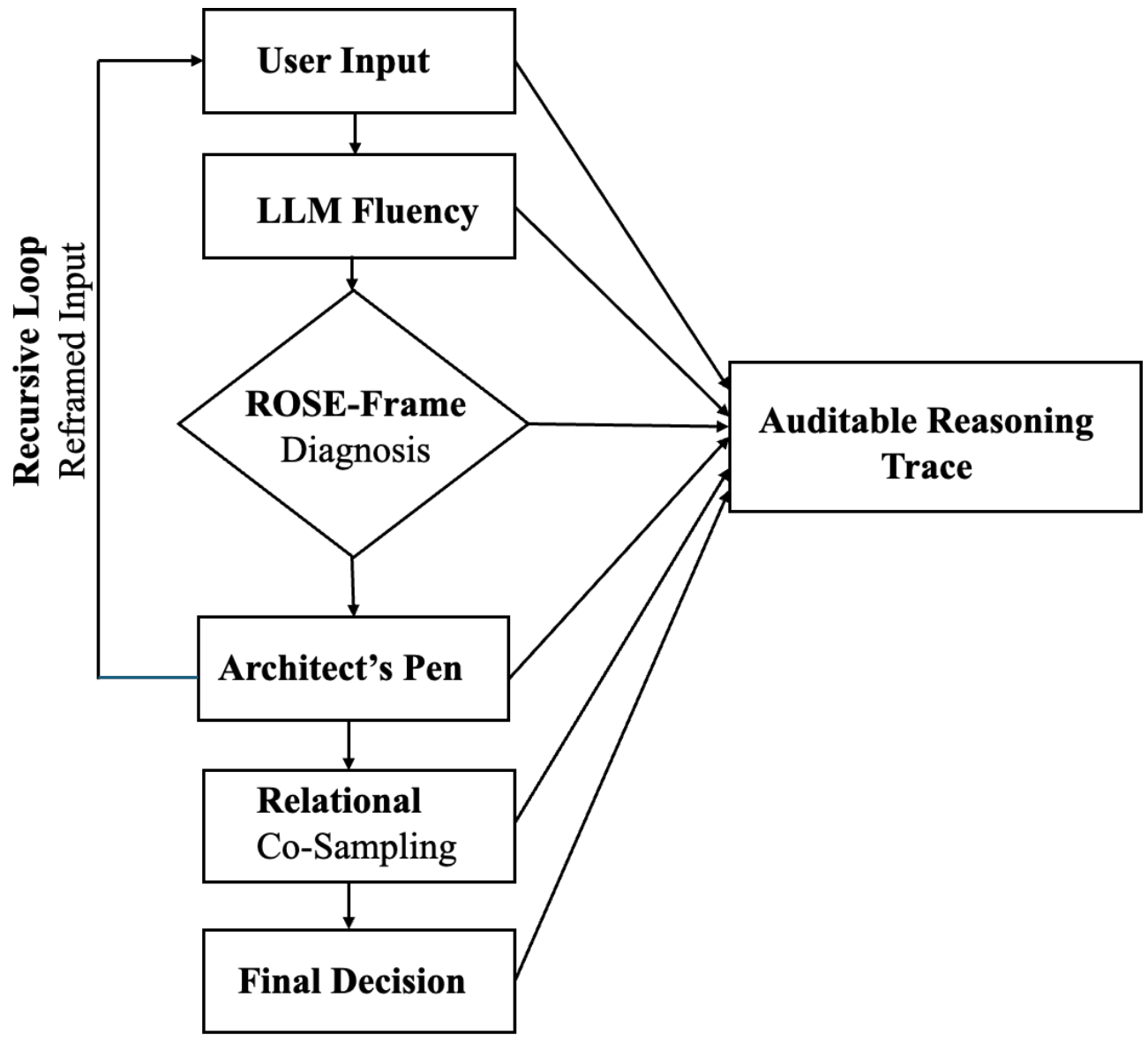

*Figure 1: RRI Architecture as an Inference-Time Middleware for Structured Human–AI Reasoning.*

The Rose-Frame diagnoses where drift is likely to occur in the interaction between humans and models. The Architect's Pen provides the mechanism for selective reflection. RRI integrates these components into an operational workflow. RRI is not a new model architecture, algorithm, or training paradigm; it is an inference-time complement that transforms human–AI interaction into a reliable reasoning system (see Figure 1).

The guiding idea is simple: the unit of intelligence is no longer the model alone, but the human–model relation. RRI ensures that this relation becomes cognitively stable, falsifiable, and auditable.

Where traditional AI systems attempt to localise "intelligence" inside the model, RRI distributes intelligence across three functions: (i) Human abstraction, (ii) Model articulation, (iii) Human reflection and correction. This triad forms a structured reasoning loop that maintains meaning under pressure, uncertainty, and complexity.

**What RRI Does**

RRI enables:

- Selective System-2 engagement when cognitive traps are detected.
- Joint reasoning where both human and model contribute unique strengths.
- Context-sensitive reflection without slowing all interactions.
- Inference-time governance that requires no model retraining.
- Traceable reasoning through Auditable Reasoning Traces (ARTs).
- Adaptation to risk level, from creative exploration to regulated decisions.

RRI is therefore the missing operational layer in human–AI collaboration, analogous to the role of operating systems in mediating between hardware and applications. It provides the cognitive "interface protocol" that makes LLM-supported reasoning safe, reflective, and institutionally accountable.

### Why Existing Approaches Don't Solve This

Current AI-safety methods focus on shaping the model rather than on stabilising the interaction where human–AI reasoning actually occurs (see Table 2). Constitutional AI constrains *what the model says*, but does not influence *how users evaluate* its outputs.[30] Chain-of-Thought increases transparency, yet it offers no structure for human reflection and can still be accepted uncritically.[31] RLHF aligns models with human preferences rather than with grounded, falsifiable reasoning.[30] Explainable AI provides post hoc explanations, but evidence increasingly indicates that these explanations often increase user overconfidence rather than correct it.[32,33]

| Approach | What It Does | What It Misses | How RRI Complements |
|---|---|---|---|
| **Bigger models (e.g., next-generation LLMs)** | Increase accuracy and coverage | Do not govern human reasoning or evaluation | RRI works with any model, regardless of size |
| **Constitutional AI[30]** | Constrains and guides model outputs | Does not structure human reflection or judgment | RRI + Constitutional AI provide a full-stack interaction architecture |
| **RAG / Grounding** | Ties answers to external sources | Citations do not ensure critical evaluation | RRI makes evaluation explicit through reflection steps |
| **Explainable AI[32,33]** | Provides explanations for outputs | Explanations can increase user overconfidence | RRI structures how explanations are interpreted and used |
| **AI Red-Teaming** | Identifies model failures under stress | Does not support everyday reasoning for typical users | RRI offers continuous, everyday governance during normal use |

*Table 2. Limitations of current AI-safety approaches and the complementary role of RRI*

RRI is not a substitute for these methods. It is a complementary layer that governs the reasoning process between human and model, ensuring that reflection, conflict, and grounded evaluation occur precisely where drift most often arises.

### The Illusion of Reasoning: Surface Simulations and "Hindsight Explanations."

A growing set of techniques, Chain-of-Thought prompting, "reasoning tokens," and long-deliberation modes, create a strong impression that large language models now perform genuine reasoning (Table 3). Yet the available evidence indicates that these mechanisms are surface-level simulations rather than cognitive processes. The model completes its latent computation in a fast, associative pass and only afterwards generates the step-by-step narrative, producing a post-hoc justification rather than a causal reasoning chain.[34–36]

A simple example illustrates this: When asked, "Why is the sky blue? Show your reasoning," the model predicts the answer ("Rayleigh scattering") immediately and then constructs an explanation to match the requested format. Changing the instruction ("give two steps," "give ten steps," "explain like a physicist") modifies the narrative but leaves the answer unchanged. The steps are written to satisfy the style, not to conclude.

Removing or altering the explanation alters human perception, but model accuracy changes little. In Kahneman's terms, these traces function as hindsight explanations rather than evidence of internal deliberation.[28] The reasoning is for humans, not for the model.

This is the relational failure predicted by the Rose-Frame: humans mistake fluency for depth, structure for deliberation, and format for cognition. The model remains a System-1-style engine, fast, associative, heuristic, while the user relaxes their own System 2 engagement because the output appears reflective. Reasoning never occurs in the model, yet the user behaves as if it has.[34]

Because this "reasoning" lives only on the linguistic surface, these approaches provide no support for conflict, falsification, or epistemic recalibration; the core functions needed to stabilise human–AI collaboration. The reflective burden silently shifts to the human, even as the human believes the opposite.

| **Apparent Innovation** | **Claimed Benefit** | **What Actually Happens** | **Rose-Frame Diagnosis** |
|---|---|---|---|
| Chain-of-Thought "reasoning tokens" | "The model thinks step-by-step." | The answer is computed first; the steps are generated as narrative decoration. | Fluency mistaken for depth. |
| Long deliberation / long context | "More tokens = more reasoning." | Longer monologue, same latent computation. | Confirmation by volume. |
| Visible "thinking mode" | "Transparent reasoning process." | Human overconfidence ↑; no improvement in grounding. | Map–territory collapse. |

*Table 3: Why Current Methods Remain Surface Simulations*

RRI intervenes at the human–AI interface, where the illusion of reasoning is strongest and where epistemic drift most frequently arises. By converting a simulated monologue into a structured, auditable interaction, RRI replaces the appearance of reasoning with the practice of reasoning. It is not a replacement for model-level safety; it is the missing scaffolding that ensures that reasoning occurs between human and model, rather than merely being performed on the screen.

**RRI Workflow Overview**

A RRI cycle unfolds in four steps:

1. Signal Detection (Rose-Frame): The interaction is monitored for signs of drift: excessive fluency, weak grounding, premature coherence, ambiguous user intent, missing alternatives or counterfactuals, and ontologically thin explanations. When a risk signal appears, RRI activates the intervention layer.

2. Intervention (Architect's Pen): The Pen introduces a targeted reflective step, either the three-step relational loop or multiple loop control cycle, depending on risk level.

3. Structured Interpretation (Human–Model Loop): The model's output is tested, clarified, falsified, or reframed. Human reasoning becomes visible and explicit.

4. Documentation (ART): If needed, the system generates a traceable record that links intent → analysis → reflection → outcome. This enables review, accountability, and compliance.

RRI iteratively repeats this cycle, refining reasoning at each pass.

**Why RRI Must Operate at Inference Time**

Inference-time[37] interventions have several advantages:

- Faster and cheaper than training-time changes
- Adaptive to context (creative vs analytical vs regulated)
- Model-agnostic (works on GPT, Claude, Gemini, Llama)
- Compatible with rapid model iteration cycles
- Immediately deployable at scale
- Creates cognitive structures where none exist

Most importantly, inference-time reflection leverages the strengths and limitations of LLMs rather than working against them. Large models excel at generating possibilities, patterns, narratives, and reformulations. They are well-suited to support human reasoning, but only if the human is guided to actively evaluate, question, and correct the output. RRI structures for the evaluation.

**What Makes RRI Distinct**

RRI operates at inference time and structures interaction with model output, but it should not be conflated with metacognitive scaffolding, human-in-the-loop oversight, chain-of-thought prompting, cognitive forcing functions, or reflective HCI systems. These approaches introduce local prompts or reflective cues, whereas RRI differs in its unit of analysis and governance role.

Traditional methods treat reflection as a content-level intervention - by adding explanations, requests for justification, or interface nudges. RRI instead treats reflection as a relational property of the human–model dyad. It governs how the human and the model coordinate their reasoning, using a stable interaction protocol triggered by drift signals and structured through the Architect's Pen.

Unlike human-in-the-loop systems, which position the human as an external reviewer, RRI casts both the human and the model as co-contributors in an iterative reasoning loop focused on internal cognitive stability rather than on output checking.

RRI is also not a prompting technique. Chain-of-thought, self-consistency prompts, or lengthy explanations simulate deliberation but leave the model's fast, associative computation unchanged. RRI counters this illusion by shifting the locus of reasoning from the model to the interaction, making the human's evaluative steps explicit and auditable.

Finally, RRI goes beyond cognitive forcing functions or reflective interface cues. It coordinates a structured reflective cycle, generates Auditable Reasoning Traces when required, and functions as an inference-time governance layer. Its innovation lies in the architecture, not the prompts: a relational system designed to stabilise reasoning where drift actually occurs, at the intersection of human intuition and model fluency.

# 6. RRI Reflection Modes: Implementation and Methods

The conceptual architecture presented in earlier sections becomes operational only when translated into concrete methods. This section outlines how RRI employs reflection in practice and how the system adapts to task difficulty, cognitive risk, and domain requirements. It introduces the reflection modes, the five-step control cycle, and the Auditable Reasoning Trace, showing how each component turns high-level principles into measurable, real-world behaviour.

## Core Mechanism: The Five-Step Reflective Control Cycle

The Architect's Pen provides the conceptual mechanism for structured reflection (developed in Part III). In this paper, we operationalise that mechanism through a five-step reflective control cycle. This cycle is the core of the RRI workflow: a lightweight, inference-time sequence that reintroduces reasoning at the precise moments where the Rose-Frame detects drift. The logic is simple: awareness → falsification → delay → accountability → feedback.

Each step corresponds to a cognitive mechanism that describes how humans detect bias, regulate impulsivity, generate counterfactuals, and learn from outcomes. The cycle does not moralise reflection; it operationalises it. The Pen intervenes only when needed, when fluency dominates, grounding weakens, or coherence suppresses conflict. The five steps are summarised in Table 4.

| Step | Cognitive Target | Purpose | Example Prompt | Explanation |
|---|---|---|---|---|
| **1<br>Cognitive State** | Affect regulation | Detect whether the user is deciding under stress, fatigue, or attraction | *"Are you making this judgment under stress, time pressure, or desire for a quick answer?"* | Emotional states often bias cognition. Surfacing the user's state reduces silent drift and prevents intuitive shortcuts from dominating. |
| **2<br>Counterfactual** | Counter-argument generation | Challenge initial assumptions and surface alternatives | *"What evidence would prove this interpretation wrong?"* | This step forces deliberate conflict, the most potent antidote to coherence bias and confirmation drift. |
| **3<br>Temporal** | Impulse control | Prevent premature closure on the first coherent narrative | *"Wait 15 minutes or revisit tomorrow, does your judgment change?"* | Even minimal delay improves accuracy by reducing noise, impulsivity, and pressure-driven decisions. |

| Step | Cognitive Target | Purpose | Example Prompt | Explanation |
|---|---|---|---|---|
| **4 Transparency** | Justification pressure | Ensure decisions withstand peer or institutional scrutiny | *"Would you stand by this reasoning if reviewed by a colleague or regulator?"* | Accountability reliably improves the quality of reasoning by shifting users from intuition to justification. |
| **5 Outcome** | Feedback-based learning | Capture the decision and expected outcome for later comparison | *"Record your expected outcome. Revisit in 30 days to compare prediction and reality."* | Closing the loop with feedback builds calibration and long-term improvement in judgment. |

*Table 4: The five-step reflective control cycle and its cognitive functions*

Each step in the control cycle corresponds to well-established mechanisms in cognitive and behavioural science. Affect awareness has been shown to improve judgment under uncertainty by reducing emotion-driven noise.[18,38,39] Counterfactual reasoning is a cornerstone of scientific method and forecasting accuracy, supporting the generation of alternative hypotheses and error detection.[29,40] Temporal delay reliably reduces impulsivity and increases decision consistency, even when the delay is minimal.[28,41] Accountability pressures shift individuals from intuitive acceptance to reasoned justification, improving accuracy and reducing bias. Finally, outcome feedback, particularly when predictions are recorded and revisited, builds calibration and long-term learning.[42] Together, these mechanisms provide a robust cognitive foundation for the five-step reflective cycle.

**Reflective Control Cycles Relation to Rose-Frame**

Each step in the reflective control cycle corresponds to one or more of the cognitive traps identified by the Rose-Frame. The alignment is structural, not incidental:

System-1 dominance is primarily countered by Steps 1, 3, and 4. The State step reduces emotion-driven shortcuts by surfacing the affective pressures that often trigger intuitive drift. The Temporal step slows premature closure, creating the pause necessary for deliberation. The Transparency step then forces justification rather than intuition, requiring the user to articulate reasons rather than accept the first coherent answer. Together, these mechanisms interrupt fast, fluent reasoning and reintroduce selective System-2 engagement.

Epistemology–ontology confusion is addressed by Steps 2 and 4. The Counterfactual step compels the user to distinguish between linguistic representation and actual evidence by asking what would disconfirm the model's output. The Transparency step reinforces this distinction by requiring grounding that withstands external scrutiny. In combination, these steps prevent the common tendency to treat text as truth or coherence as evidence.

Confirmation-driven convergence is mitigated by Steps 2 and 5. The Counterfactual step introduces structured conflict in the moment, forcing the user to consider alternative interpretations. The Outcome step embeds falsification across time by requiring predictions to be compared with later reality. These mechanisms break coherence loops, restore falsifiability, and ensure that agreement arises from evaluation rather than from linguistic fluency.

In low-stakes, everyday use, the Pen does not trigger this entire sequence. However, in high-stakes fields such as medicine, finance, law, education, and public governance, the complete five-step cycle may be indispensable.

## Output Layer: The Auditable Reasoning Trace (ART)

Reflection alone is insufficient for institutional reliability. High-stakes domains require traceability, the ability to reconstruct how a decision was reached. The Architect's Pen, therefore, generates an Auditable Reasoning Trace (ART)**,** a lightweight, timestamped record that links:

1. User framing
2. Model output
3. Reflection steps
4. Final judgment

An ART functions as a cognitive equivalent of a clinical note, legal memorandum, or audit trail. It is not surveillance; it is documentation. ARTs enable: transparent review, accountability, detection of shallow reasoning or click-through, comparison of reasoning quality across cases, and regulatory compliance.

Importantly, ARTs can be stored locally or at the institution. They require no model retraining. They maintain privacy while enabling scrutiny. And they satisfy emerging requirements under the EU AI Act, ISO 42001, and sector-specific compliance frameworks. By combining the diagnostic power of the Rose-Frame with the operational precision of the Architect's Pen, ARTs close the loop between cognition, governance, and accountability.

## Customisation Layer: Reflection Modes Related to Context

Different tasks demand different levels of reflection. The strength of RRI lies in its ability to adjust the intensity of structured reasoning based on domain, stakes, and cognitive load. Table 5 summarises our four suggested reflection modes.

### The Open Pen: Low-Stakes, High-Volume Use

In everyday interactions, students drafting ideas, consumers asking questions, or writers exploring themes and reasoning are predominantly intuitive and fast. Users have limited tolerance for delay, limited awareness of reasoning structure, and little need for documentation. In this environment, RRI deploys only minimal cues, such as offering an alternative viewpoint or highlighting an assumption. The goal is not to impose structure but to subtly steer cognition toward slightly more reflective habits without interrupting flow. Even light scaffolding at the population scale can shift public reasoning norms over time.

| Mode | Cognitive Profile | RRI Behaviour | Typical Domains |
|---|---|---|---|
| **The Open Pen** | Fast, intuitive, low-friction reasoning | Light cues, no logging, privacy-first | Everyday tasks, consumers use |
| **The Creative Pen** | Imaginative and exploratory thinking | Optional reflection, narrative reframing | Creative writing, concept design |
| **The Guided Pen** | Semi-structured knowledge work | Intermediate reflection and clarification | Strategy, education, analysis |
| **The Accountable Pen** | High-stakes, regulated reasoning | Full reflective cycle + ART | Medicine, law, finance, governance |

*Table 5: Customisation Layers*

### The Creative Pen: Creative and Exploratory Use

Creative work requires a different cognitive posture. When users generate metaphors, stories, speculative designs, or conceptual sketches, the relevant dimension is expansion rather than validation. RRI largely steps back, allowing the LLM to expand its imaginative range without enforcing an argumentative structure. Reflection remains optional and context-driven, a prompt to challenge clichés, invert perspectives, or explore alternatives when the user signals interest. Here, RRI functions as a flexible companion to creativity rather than as a governance layer.

### The Guided Pen: Knowledge Work and Education

In semi-structured domains such as strategy analysis, academic work, policy writing, or market research, reasoning blends intuition and deliberate structure. Users face moderate cognitive risk and often need reproducible reasoning steps. RRI introduces intermediate-level reflection: counterfactual prompts, justification checks, and clarification of assumptions. Documentation remains optional but beneficial, enabling teachers, supervisors, and managers to evaluate reasoning rather than merely final outputs. This mode strengthens analytical quality while preserving workflow fluidity.

### The Accountable Pen: Regulated and High-Stakes Domains

In regulated, high-stakes environments, such as medicine, finance, law, scientific research, compliance, and public governance, the cost of error is high, and tolerance for ambiguity is low. Decisions in these domains must withstand external scrutiny, institutional review, and, at times, legal challenge. For this reason, reflection is not optional but structurally mandated. RRI therefore activates the full five-step reflective cycle, ensuring that reasoning is deliberate, traceable, and capable of surviving adversarial evaluation. This mode surfaces uncertainty, distinguishes evidence from interpretation, and prevents premature closure on fluent but unsupported conclusions. It transforms human–AI interaction from fast conversational exchange into accountable epistemic work.

In this setting, RRI generates an Auditable Reasoning Trace for every high-stakes decision. An ART captures the user's framing, the model's contribution, the reflective steps taken, and the final judgment. This ensures justification pressure: the reasoning must be coherent enough that the user would stand by it before colleagues, regulators, or a review board. The system also detects shallow engagement, such as rapid "click-through" behaviour, which often signals uncritical acceptance of the first coherent answer. By preserving the intermediate reasoning steps, ARTs make it possible to identify when uncertainty was ignored, when conflict was suppressed, or when the model's suggestions were accepted without adequate evaluation. This prevents Goodhart-style[43] metric gaming, where users or organisations optimise for speed or volume rather than correctness and reasoning quality.

In practice, the Accountable Pen becomes a cognitive audit trail. It structures reasoning in complex clinical diagnosis, legal precedent evaluation, investment memo preparation, regulatory risk assessment, scientific interpretation, and the drafting of expert opinions. The combination of full-cycle reflection and ART documentation produces institutional reliability: decisions become reviewable, interpretable, and falsifiable. In these domains, RRI does not merely assist users; it serves as the foundation of responsible judgment under uncertainty, ensuring that human–AI collaboration meets the professional, ethical, and legal standards required for real-world deployment. The following section, therefore, examines how these mechanisms align with emerging regulatory expectations and how structured reflection can serve as a practical bridge between cognition and governance.

## 7. From Reflection to Regulation: Preparing for Compliance

We now move from interaction design to institutional oversight. The same reflective structures that stabilise reasoning at the cognitive level also satisfy emerging regulatory requirements. Governments and regulatory bodies are converging on a clear expectation: high-risk AI systems must be traceable, transparent, and governed through documented human oversight. The leading international frameworks (the EU AI Act, the OECD AI Principles, the NIST AI Risk Management Framework, and ISO/IEC 42001) share a common foundation.[13,14,44,45] They require organisations to demonstrate how decisions are reached, who exercised judgment, which uncertainties were considered, and how risks were evaluated and controlled. These obligations are particularly stringent for high-risk applications and for general-purpose AI models approaching systemic thresholds. In this context, reflection is not merely a cognitive safeguard; it becomes a mechanism of compliance.

This is consistent with the *International AI Safety Report 2026*, which emphasises that many safety interventions fail when oversight is applied to degraded or unreliable signals.[46] RRI addresses this by treating interaction stability and traceable reasoning as prerequisites for effective governance, not optional usability features.

The NIST AI Risk Management Framework is particularly explicit on this point: it defines traceability and documentation as core functions of trustworthy AI, requiring organisations to maintain structured records of how human judgment interacts with model output. ART potentially fulfils this requirement by creating a transparent, inference-time record of the reasoning process.[45]

At present, institutions using LLMs in medicine, finance, law, and the sciences often cannot meet these requirements. Interactions leave no reasoning trail capable of reconstruction, making it difficult to demonstrate "meaningful human control" (EU AI Act, Art. 14) or to satisfy the traceability and documentation obligations embedded in ISO 42001 and NIST RMF. The gap is structural: LLMs produce fluent output, but they do not record the cognitive process that led to a conclusion, nor the human reflections that shaped it.

RRI addresses these challenges by embedding governance into the reasoning process itself. Its triadic structure (WHY, WHAT, HOW) maps naturally onto the core principles of the EU AI Act. WHY clarifies the system's purpose and risk classification, ensuring that every interaction begins with explicit intent. This satisfies the Act's requirement for meaningful human control and purpose-driven design. WHAT captures the architecture and data provenance, supported by RRI's diagnostic tools that make reasoning visible and auditable. HOW operationalises oversight through inference-time reflection, and ART creates a timestamped record of framing, model outputs, reflective steps, conflicts encountered, uncertainties noted, and the final justification.

The EU AI Act sets milestones that trigger enhanced obligations depending on a model's scale and potential societal impact. Providers of General-Purpose AI (GPAI), defined as models exceeding ~$10^{23}$ FLOPs and capable of multi-domain functionality, must provide in-depth technical documentation.[47] GPAI models deemed to carry systemic risk, typically those exceeding ~$10^{25}$ FLOPs or flagged by coordinated assessment, face additional controls, including comprehensive risk evaluations and mitigation strategies, adversarial red-teaming, serious-incident reporting, and robust cybersecurity protections.[48] RRI provides a scalable approach to meeting these obligations for documentation, risk management, and human oversight. By detecting cognitive drift and enforcing structured reflection, RRI complements technical safeguards with a governance layer that is both practical and adaptive. It transforms compliance from an external burden into an internal capability, one that enhances the quality of reasoning while satisfying regulatory expectations. Regulators receive what they have repeatedly emphasised they need: demonstrable human supervision, explicit reasoning, and a transparent link between input, deliberation, and outcome.

The regulatory advantage becomes a cognitive advantage. Institutions that adopt RRI do not merely comply; they potentially cultivate higher-quality reasoning, fewer errors, and more consistent decision-making across teams. RRI transforms governance from an external constraint into an internal capability: a way of thinking that strengthens professional judgment and produces a durable epistemic record suitable for review, audit, learning, and continuous improvement.

# 8. Discussion

This paper proposes a shift in how LLMs are understood, used, and governed. Rather than treating AI as an autonomous reasoning agent, the framework recognises that reasoning emerges from the interaction between the human and the model. The Rose-Frame identifies how drift arises in this joint space. The Architect's Pen introduces a structured, reflective mechanism to regulate it. Relational Reflective Intelligence integrates both into a practical, inference-time workflow that improves reasoning without retraining models or slowing everyday tasks. The approach reframes several debates in contemporary AI research:

Hallucinations as a relational phenomenon: LLM hallucinations are often described as model errors, but this misses the core problem: hallucinations arise from the structure of human language and are amplified when humans rely on fluency as a proxy for truth. RRI shows that hallucinations are best understood and mitigated within the human–model relation.

Alignment as cognitive interaction, not autonomous control: Alignment research focuses heavily on controlling model behaviour. But most real-world risks today stem not from autonomous agents but from human misuse, misunderstanding, and overconfidence in LLM output. RRI positions alignment as a shared cognitive responsibility, achieved through reflective interaction rather than top-down behavioural constraints.

Governance as reasoning architecture: Traditional approaches to AI governance emphasise documentation, risk classification, and oversight protocols. RRI suggests that governance can also be cognitive: structured reflection is a form of oversight. ARTs allow organisations to comply with regulations because reflection is embedded into the reasoning workflow itself.

Intelligence as distributed, not centralised: RRI shifts intelligence from a property of a single system to an emergent property of well-structured interactions. This fosters a more realistic and grounded understanding of "super-intelligence": not a runaway machine, but a high-functioning human–AI dyad capable of achieving clarity and reliability in the face of uncertainty.

## Limitations

Although the framework is intended to be practical and widely applicable, it has several limitations that must be acknowledged.

The dependency on user engagement: RRI requires users to reflect at key moments. While the system minimises cognitive load by activating only when drift is detected, some users may still bypass reflective steps. This creates variability in reasoning quality. Institutional norms and incentives will influence the consistency with which RRI is applied.

Empirical validation across domains is needed: The framework synthesises established cognitive science with AI interaction design, but large-scale empirical testing remains necessary. Different fields, such as medicine, law, finance, and engineering, may require customised versions of the reflection cycle. The strength, timing, and frequency of interventions will need to be calibrated.

Implementation constraints in real-world systems: While RRI is model-agnostic and inference-only, integrating it into existing platforms requires interface design, logging infrastructure, and domain-specific adaptation. Organisations may resist incorporating reflective friction, even when the long-term benefits are clear. Early adopters will likely be in high-stakes sectors with strong governance cultures.

ARTs require careful design for privacy and storage: Auditable Reasoning Traces introduce transparency but also raise new questions about data retention, privacy, and access control. Storing cognitive traces must be done in a way that protects users while enabling oversight. Local storage, encrypted logs, and institution-specific policies will be critical.

The framework does not solve all forms of epistemic drift: RRI targets cognitive drift arising from interaction patterns and reasoning shortcuts. It does not solve domain-specific knowledge gaps or misinformation embedded in training data. It complements, but does not replace, technical advances such as retrieval-augmented generation and uncertainty estimation.

These limitations are not weaknesses of the concept, but natural boundaries that indicate where further research and adaptation are needed.

## Future Work: A Collaborative Roadmap for Reflective AI

Advancing Relational Reflective Intelligence requires coordinated work across cognitive science, AI engineering, safety research, and real-world practice. The framework cannot mature in isolation; it depends on collaboration among model developers, AI labs, enterprises, and domain practitioners. Key areas include:

**Quantifying relational drift**: We invite interdisciplinary partners to develop empirical methods to detect when human–LLM interaction deviates from grounded reasoning. Metrics for conflict detection, grounding strength, and reflective depth are needed to establish standardised benchmarks for reasoning quality across settings.

**Adaptive reflection mechanisms**: Reflection must adapt to task demands, cognitive load, user expertise, and domain risk. Collaboration with model developers and HCI researchers is essential to the design and validation of dynamic thresholds that efficiently regulate reflection during inference.

**Longitudinal studies of reasoning behaviour**: RRI generates outcome-based feedback suitable for studying how reasoning evolves. Enterprises and institutions are needed to support long-term field studies that examine how structured reflection influences accuracy, over-trust, and professional judgment in real-world operations.

**Domain-specific implementations**: Clinical practice, financial regulation, legal reasoning, and scientific publishing each have distinct cognitive structures. We seek to collaborate with domain experts and industry partners to tailor RRI to these environments, identify the most effective reflective steps, and determine how ARTs should be configured for operational workflows.

**Integration with technical AI-safety mechanisms**: RRI operates entirely at inference time, governing how humans and models reason together rather than how the model is trained. But it is designed to complement, not replace, technical safety methods such as training-time alignment, uncertainty quantification, retrieval-augmented grounding, and agent-level control architectures. The next step is collaborative research with AI labs and safety groups to understand how these two layers interact: algorithmic safeguards stabilise the model, while cognitive safeguards stabilise the interaction. When combined, they create a more robust safety envelope than either layer can provide on its own.

**Organisational and cultural impact**: Real-world deployment depends on understanding how RRI shapes institutional reasoning, accountability, and decision-making. We invite enterprises and public organisations to run pilot programs to evaluate how ARTs support quality improvement, reduce errors, and strengthen reflective practice.

RRI is therefore not a standalone system but a collaborative roadmap, one that requires joint development across disciplines to build safer, more reflective human–AI ecosystems. To this end, we propose a series of targeted pilot studies in domains where reasoning quality and regulatory compliance are non-negotiable.

The first could focus on clinical decision-making and the embedding of RRI within diagnostic workflows. Here, the goal would be to determine whether structured reflection can reduce cognitive drift and mitigate false confirmation bias. One could measure whether clinicians accept fewer plausible-but-false outputs, whether confidence calibration improves - aligning perceived certainty with actual accuracy, and whether ART can generate traceable reasoning trails that satisfy compliance requirements under the EU AI Act (Article 14) and ISO/IEC 42001. Key metrics would include reductions in false confirmation rates, stronger correlations between confidence and accuracy, and the completeness and audit readiness of ART documentation.

Another pilot could examine the legal review of contracts, integrating RRI into contract analysis platforms. Such a study would assess whether reflective scaffolding, through counterfactual prompts and justification steps, reduces premature agreement and strengthens interpretive accuracy. It would also validate ART as a mechanism for traceability and accountability under frameworks such as the NIST AI Risk Management Framework and the OECD AI Principles. Metrics would include the number of counterfactuals generated, the quality of the justifications provided, and a compliance traceability score that reflects the robustness of the reasoning records.

Together, these pilots could provide empirical evidence of RRI's dual benefits. Cognitively, we expect improved reasoning quality through structured reflection and falsification. In terms of regulation, we anticipate demonstrable compliance with emerging AI governance standards, achieved through auditable reasoning trails that enable practical, transparent oversight.

We invite healthcare institutions, law firms, and compliance bodies to collaborate in these pilots. By partnering in this effort, stakeholders can help establish benchmarks for reflective

AI governance and set a new standard for transparency in reasoning in high-stakes environments.

## 9. Conclusion

LLMs have created a world in which information is abundant, but reflection is scarce. The central challenge is not hallucination or a lack of reasoning within the model; it is the erosion of reasoning in the interaction. Humans increasingly outsource judgment to fluent outputs, reinforcing shortcuts that both humans and models share. Without new structures, this mutual drift becomes invisible, predictable, and potentially harmful in high-stakes environments.

This paper argues that the solution is neither to slow down AI progress nor to impose heavy-handed restrictions. Instead, reasoning itself must be restored as the core of human–AI collaboration. The Rose-Frame reveals where reasoning drifts. The Architect's Pen reintroduces structure and clarity to the workflow. Relational Reflective Intelligence operationalises reflection, transforming it into a scalable cognitive infrastructure.

Where *Attention Is All You Need*[15] enabled models to construct meaning through relations between words; this framework allows humans and machines to create understanding through mutual relations. Attention built the model's fluency; reflection builds the system's reliability. The future of intelligence is not a contest between human and machine, but a partnership, a distributed cognitive system that is more stable, transparent, and grounded than either agent alone.

In this sense, RRI not only mitigate risk; it establishes the conditions for a new form of epistemic stability. Its relational layer builds on themes already emerging in the work of leading AI theorists. Bengio's call for System-2–like inductive biases highlights the need for deliberate, decomposable reasoning inside the model;[49,50] Hassabis's emphasis on world-model architectures underscores the importance of structured understanding capable of simulation.[9] RRI extends these ideas into the interaction space itself, the human–AI relation, acknowledging that current System-2-like advances remain simulations. Rather than seeking reasoning inside the model, RRI embeds governance externally, making reflection a practical, auditable process. As noted previously, this perspective aligns with emerging themes in AI research, where scholars such as Bengio and Hassabis emphasise structured reasoning and world models as prerequisites for safe intelligence.

We argue that this interaction-level layer is the missing complement: the 'System 3' that coordinates human reflection with model inference. It turns consumption into reflection, fluency into understanding, and dialogue into a site of structured learning. Under this view, the next phase of AI progress is not autonomous super-intelligence but Relational Reflective Intelligence, where reasoning is stabilised not by the machine alone but by the architecture of the relation.